\documentclass[seceq]{ptptex}

\usepackage{graphicx}
\usepackage{wrapft}

\notypesetlogo                       %comment in if to eliminate PTPTeX
\title{%        %You can use \\ for explicit line-break
Quantum Manifestations of Classical Stochasticity\\ in the Mixed
State }

\author{%       %Use \scshape  for the family name
Victor~P.~\textsc{Berezovoj}, Yuri~L.~\textsc{Bolotin},
Vitaliy~A.~\textsc{Cherkaskiy} }

\inst{%         %Affiliation, neglected when [addenda] or [errata]
Institute for Theoretical Physics\\
National Science Center ``Kharkov Institute of Physics and Technology''\\
61108 Kharkov, Ukraine}

\abst{%         %this abstract is neglected when [addenda] or [errata]
We investigate the QMCS in structure of the eigenfunctions,
corresponding to mixed type classical dynamics in smooth potential
of the surface quadrupole oscillations of a charged liquid drop.
Regions of different regimes of classical motion are strictly
separated in the configuration space, allowing direct observation
of the correlations between the wave function structure and type
of the classical motion by comparison of the parts of the
eigenfunction, corresponding to different local minima.}

\begin{document}

\maketitle

\section{Introduction}
The deformation potential
\begin{equation}
\label{1} U(a_0,a_2)=\sum_{m,n}C_{mn}(a_0^2+2a_2^2)^m a_0^n
(6a_2^2-a_0^2)^n
\end{equation}
where $a_0$ and $a_2$ are the internal coordinates of the drop
surface
\begin{equation}
R(\theta,\varphi)=R_0\{1+a_0 Y_{2,0}(\theta,\varphi)+a_2
[Y_{2,2}(\theta,\varphi)+Y_{2,-2}(\theta,\varphi)]\}
\end{equation}
\begin{wrapfigure}{l}{\halftext}
%\figurebox{\halftext}{\halftext}
\includegraphics[width=\halftext,height=\halftext]{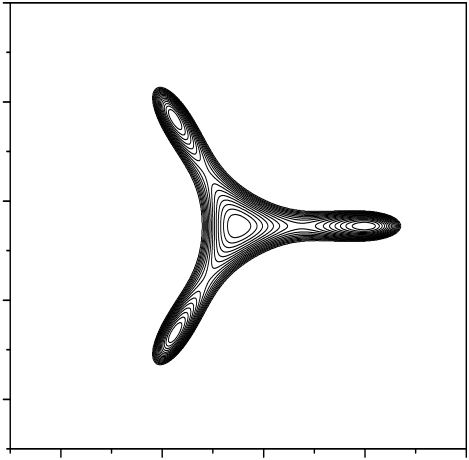}
\caption{The potential shape for $W=18$.}
\end{wrapfigure}
describes surface quadrupole oscillations of a charged liquid drop
of any nature, including atomic nuclei\cite{r26} and metal
clusters,\cite{p19} containing specific character of the
interaction only in the coefficients $C_{mn}$. Expanding (\ref{1})
to fourth order in deformation variables
\begin{equation}
x = \frac{\sqrt{2}C_{20}}{3C_{01}}a_0, y =
\frac{\sqrt{2}C_{20}}{3C_{01}}a_2
\end{equation}
and assuming equality of masses for the two independent
directions, we get the one-parametric $C_{3v}$-symmetric
Hamiltonian
\begin{equation}
\label{2} H = \frac{p_x^2+p_y^2}{2m}+U(x,y,W)
\end{equation}
\begin{equation}
\label{3} U(x,y,W) = \frac{1}{2W}(x^2+y^2)+xy^2-\frac 1 3 x^3
+(x^2+y^2)^2
\end{equation}
where $W=9C_{01}^2/(C_{10}C_{20})$.\\ The potential (\ref{3}) is a
generalization of the well-known Henon-Heiles potential\cite{2}
with one important difference: motion in (\ref{3}) is finite for
all energies, assuring existence of the stationary states in
quantum case. For $W>16$ the potential energy surface has seven
critical points: four minima (one central and three peripheral)
and three saddles. We consider in detail the case $W=18$, when all
four minima have the same depth $E=0$ and the saddle energies are
$E_S=1/20736$. Critical energy of transition to chaos equals $E_S$
for the peripheral and roughly $E_S/2$ for the central minimum, so
we will be interested in the energy range $E_S/2<E<E_S$, where the
classical motion is chaotic in the central and purely regular in
peripheral minima, resulting in what we shall call the mixed
state.\cite{p14,1}\\ We find numerically the spectrum $E_n$ and
the eigenfunctions $\psi_n(x,y)$ for the Hamiltonian (\ref{2}) by
the spectral method,\cite{3,r17} which implies the numerical
solution of the time-dependent Schr\"odinger equation
\begin{equation}
i\frac{\partial\psi(x,y,t)}{\partial
t}=\left[-\frac{\partial_x^2+\partial_y^2}{2m}+U(x,y,W)\right]\psi(x,y,t)
\end{equation}
with the symmetrically split operator algorithm\cite{sm14}
\begin{equation}
\psi(x,y,t+\Delta t)=e^{i\frac{\Delta t\nabla^2}{4m}}e^{-i\Delta t
U(x,y,W)}e^{i\frac{\Delta t\nabla^2}{4m}}\psi(x,y,t)+O(\Delta t^3)
\end{equation}
where $\exp{(i\Delta t\nabla^2/4m)}\psi(x,y,t)$ is efficiently
calculated using the fast Fourier transform. Initial wave function
\begin{equation}
\label{psi0} \psi(x,y,t=0)=\sum_na_n\psi_n(x,y)
\end{equation}
is chosen to assure the convergence of $\psi(x,y,t)$ in both the
coordinate and reciprocal spaces and to avoid the degenerate
states in the decomposition (\ref{psi0}).\\ Having calculated
$\psi(x,y,t)$ for $0<t<T$, we obtain the spectrum $E_n$ as the
local maxima of
\begin{equation}
P(E)=\frac 1 T
\int\limits_0^Tdte^{iEt}P(t)w(t)=\sum_n|a_n|^2\delta_T(E-E_n)
\end{equation}
and the eigenfunctions $\psi_n(x,y)$ as
\begin{equation}
\psi_n(x,y)=\frac 1 T\int\limits_0^Tdt\psi(x,y,t)w(t)e^{iE_nt}
\end{equation}
where
\begin{eqnarray}
P(t)=\int dxdy\psi^*(x,y,t=0)\psi(x,y,t)\\
\delta_T(E-E_n)=\frac 1 T\int\limits_0^Tdtw(t)e^{i(E-E_n)t}
\end{eqnarray}
and $w(t)=1-\cos(2\pi t/T)$ is the Hanning window function.
\section{QMCS in the eigenfunction structure}
\begin{figure}
\parbox{\halftext}{
%\figurebox{\halftext}{\halftext}
\includegraphics[width=\halftext,height=\halftext]{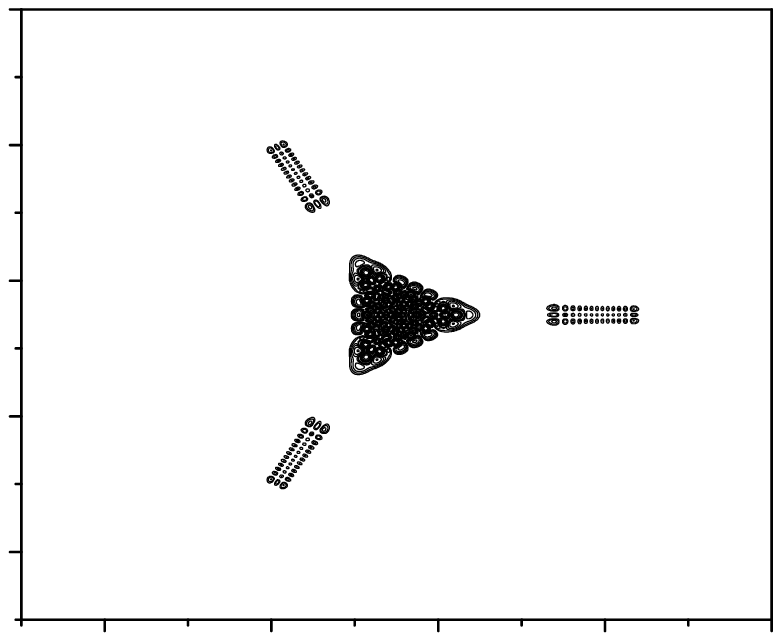}
\caption{The eigenfunction shape\label{abs}}}
\parbox{\halftext}{
%\figurebox{\halftext}{\halftext}
\includegraphics[width=\halftext,height=\halftext]{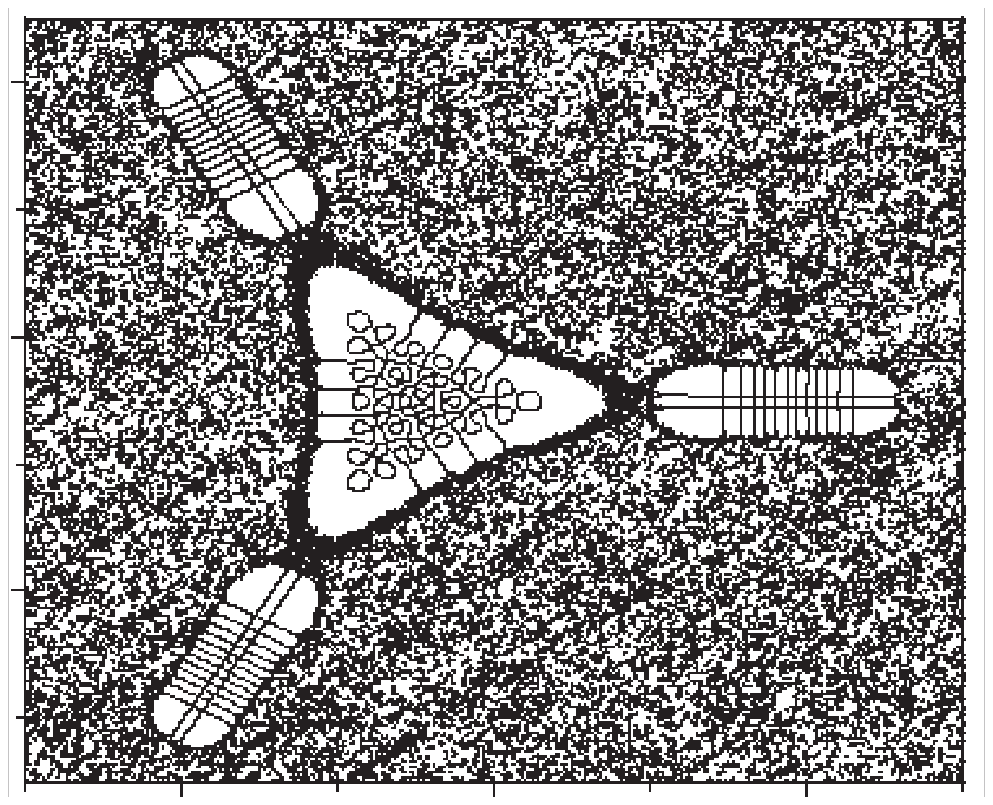}
\caption{The nodal curves\label{nod}}}
\end{figure}

Quantum manifestations of classical stochasticity can be expected
in the form of some peculiarities of concrete stationary
state\cite{cv26}\tocite{cv60} or in the whole group of states
close in energy.\cite{cv17}\tocite{cv89} Of course, it is not
excepted that such alternative does not exist at all, i.e. the
manifestations of the classical chaos can be observed both in the
properties of separate states and in their sets. For example,
comparing the eigenfunctions, corresponding to energy levels below
the critical energy of transition to chaos $E_c$, with those
corresponding to energy levels above $E_c$, drastic changes in the
eigenfunction structure can be easily seen and analyzed. However,
a question can arise if those changes are due to change in the
type of classical motion or they are just a result of the quantum
numbers change. Instead of comparing the eigenfunctions
corresponding to different states, we propose to look for QMCS in
the single quantum mechanical object --- the eigenfunction of the
mixed state.\\
$C_{3v}$-symmetric Hamiltonian (\ref{2}) is invariant under
rotation of $2\pi/3$ in the $(x,y)$ plain and under reflection
through $x$ axis, so according to the transformation properties
\begin{eqnarray}
\psi(r,\varphi+2\pi/3)=R\psi(r,\varphi)\\
\psi(r,-\varphi)=\sigma\psi(r,\varphi)
\end{eqnarray}
the eigenfunctions can be divided into three types of different
symmetry: $A_1(R=1,\sigma=1)$, $A_2(R=1,\sigma=-1)$, and double
degenerate $E(R=\exp{(\pm2\pi/3)},\sigma=\pm 1)$. The initial wave
function (\ref{psi0}) was chosen to excite only $A_1$-type states,
and we computed the energy levels and the eigenfunctions for
(\ref{2}) with $W=18$ and $m=10^{10}$, which corresponds to the
main quantum numbers of order $10^2$ in the energy range
$E_S/2<E<E_S$. Computations were made on a $1024\times1024$ grid
of length $0.6$ (distance from the central minimum to saddles was
$1/12$ and $1/6$ to the peripheral minima),
number of time steps was 16384 with increment $\Delta t = 10^4$.\\
Isolines of probability density and nodal curves for the
eigenfunction corresponding to $E_n=0.444\times10^{-4}=0.92E_S$
are presented in Fig.\ref{abs},\ref{nod}.
As we did not compute the entire spectrum, we cannot point out the
number $n$ exactly, but we estimate it to be about 200.
Correlations between the character of classical motion ---
developed chaos in the central minimum and pure regularity in the
peripheral ones --- can be easily seen comparing the parts of the
eigenfunction corresponding to different local minima.
\section{Conclusion}
We considered the quantum manifestations of classical
stochasticity in the structure of eigenfunction in the coordinate
representation, corresponding to the mixed state in the potential
of surface quadrupole oscillations of a charged liquid drop.
Correlations between character of classical motion and structure
of the eigenfunction parts, corresponding to local minima with
different type of classical motion, was observed in the shape of
the probability density distribution and in the nodal curves
structure, which gives direct and natural way to study the QMCS in
smooth potentials. Numerical computations were made by the
spectral method, which is a promising alternative to the matrix
diagonalization method for the potentials with few local minima.
%\section*{Acknowledgements}
%We would like to thank ...........

%\appendix
%\section{First Appendix} %Empty argument \section{} yields `Appendix'.
%
%\section{Second Appendix}

\end{document}